\documentclass[conference]{IEEEtran}
\IEEEoverridecommandlockouts
\usepackage{cite}
\usepackage{amsmath,amssymb,amsfonts}
\usepackage{algorithmic}
\usepackage{graphicx}
\usepackage{textcomp}
\usepackage{xcolor}
\usepackage{array}
\usepackage{multirow}
\def\BibTeX{{\rm B\kern-.05em{\sc i\kern-.025em b}\kern-.08em
    T\kern-.1667em\lower.7ex\hbox{E}\kern-.125emX}}
\begin{document}

\title{Multi-Objective Optimization for Energy- and Spectral-Efficiency Tradeoff in In-band Full-Duplex (IBFD) Communication\\
}
\author{Ata Khalili$^{\dag}$,~Sheyda Zarandi$^{\dag}$,~Mehdi Rasti$^{\dag}$, and Ekram Hossain$^{*}$\\
$^{\dag}$~Department of Computer Engineering and IT, Amirkabir University of Technology, Tehran, Iran\\
$^{*}$Department of Electrical and Computer Engineering, University of Manitoba, Canada\\
Email:~ata.khalili@ieee.org,~sheydazarandi@aut.ac.ir,~rasti@aut.ac.ir, Ekram.Hossain@umanitoba.ca}
\maketitle

\begin{abstract}
The problem of joint power and sub-channel allocation to maximize energy efficiency (EE) and spectral efficiency~(SE) simultaneously~in in-band full-duplex (IBFD) orthogonal frequency-division multiple access (OFDMA) network is addressed considering users' QoS  in both uplink and downlink.~The resulting optimization problem is a non-convex mixed integer non-linear program~(MINLP) which is generally difficult to solve. In order to strike a balance between the EE and SE, we restate this problem as a multi-objective optimization problem (MOOP) which aims at maximizing system's throughput and minimizing system's power consumption, simultaneously. To this end, the $\epsilon$-constraint method is adopted to transform the MOOP into single objective optimization problem (SOOP).~The underlying problem is solved via an efficient solution based on the majorization minimization (MM) approach.~Furthermore, in order to handle binary subchannel allocation variable constraints, a penalty function is introduced.~Simulation results unveil interesting tradeoffs between EE and SE.
\end{abstract}

\begin{IEEEkeywords}
Full-duplex (FD) communication, energy-efficiency~(EE),~spectral-efficiency~(SE),~mixed integer non-linear program~(MINLP),~multi-objective optimization problem (MOOP),~$\epsilon$-method,~majorization minimization (MM).
\end{IEEEkeywords}

\section{Introduction}
 
Efficient allocation of radio resources is necessary for both improving users' satisfaction and decreasing operators' expenditures. In-band full-duplex (IBFD) communications is a promising technology to improve both spectral efficiency (SE) and energy efficiency (EE)  in cellular wireless networks~\cite{2,3}. In IBFD communications, the ability to send and receive data simultaneously in one frequency band, can almost double the spectrum efficiency and may be exploited for reducing systems' total power consumption. Nevertheless, as the result of increase in frequency reuse factor, intensified interference, specially self-interference (SI), is a major challenge in IBFD communications. Thus, interference management achieved through precise control of network resources plays a key role in improving SE and EE when IBFD communications are employed.

There are a plethora of literature that are focused on IBFD communications in cellular networks. In many of these works such as \cite{4,5,6,7,8,9,10}, system
throughput maximization is investigated while in some others, for instance \cite{12} and \cite{13}, network energy consumption
minimization is addressed. The problem of joint subchannel and power allocation in a network with one full-duplex base station (BS) and multiple half-duplex (HD) user equipment is considered in \cite{4,5,6}. In \cite{4}, after relaxing the binary subchannel allocation variables into continuous ones, an iterative resource allocation algorithm is developed. The algorithm proposed in \cite{5} is based on decomposition and power control is addressed only after subchannel allocation policy  using a heuristic approach. In the iterative algorithm proposed in \cite{6}, subchannel assignment is determined using gradient method and power allocation is obtained after deriving a lower bound for the rate functions. In \cite{7},~subchannel assignment, power control, and duplexing mode selection are addressed using two heuristic algorithms, while in \cite{8}, only the problem of power allocation is addressed when both SI and cross-tier interference are taken into account. Furthermore, the authors in \cite{9} investigate the resource allocation algorithm for multicarrier non orthogonal multiple access systems employing a FD-BS and HD users.~Then,~a monotonic optimization is employed to find the joint power and subchannel
in order to maximize the network throughput. In \cite{11},~resource allocation schemes are proposed for EE maximization in the downlink (DL) of orthogonal frequency division multiple access (OFDMA) cellular networks with energy harvesting capability is proposed where the alternating direction method of multipliers (ADMM) and fractional
programming. The resource allocation for multiuser network incorporating full-duplex multi-antenna BSs is studied in \cite{13}, where the objective is to minimize the total
power consumption through jointly optimizing the downlink beamformer, uplink transmit power and antenna selection at BSs. 

Multi-objective optimization has also been studied in previous literature in order to strike a balance between the considered competing objectives. In \cite{14}, resource allocation
for obtaining the tradeoff between EE and SE in a single-link network is addressed.  In the absence of interference, since the rate function is convex, the tradeoff between the aforementioned objective functions, is achieved through a simple algorithm, using the weighted sum method.~The tradeoff between EE and SE in an FD network where the users operate in HD mode is investigated for two models of residual self interference~(RSI),~namely,~the constant~RSI and the linear RSI model~\cite{15}. 
In \cite{16}, IBFD communications in a single cell network with FD BS and HD users is considered. The goal of the modeled multi-objective optimization problem (MOOP) is to derive a trade-off between minimizing DL and uplink (UL) transmit power and maximizing harvested energy. This problem is then optimally solved using semi-definite program
relaxation.

In the above context, the contributions of our paper can be summarized as follows:
		\begin{itemize}
		\item We investigate the problem of joint subchannel and power allocation to strike a balance between EE and SE in an OFDMA network.~This is in contrast with other existing literature such as \cite{4}-\cite{10}, in which the objective function is system throughput maximization or \cite{13} which focuses, solely, on minimizing system's total power consumption.
		
		\item Contrary to \cite{8},  in this paper, we take both subchannel and power allocation problems into account and unlike the proposed algorithms in \cite{5,6,7}, we jointly optimize the aforementioned variables and obtain the resource allocation scheme in a single step.
		\item We take users' QoS into account and a minimum data rate is guaranteed for each user both in uplink (UL) and DL. This is unlike \cite{4}-\cite{7},~and \cite{9}~in which users' QoS requirement is completely neglected.
	\end{itemize}


\section{System Model and Assumptions}
We consider an IBFD OFDMA network with one macro base station (MBS) and $N$ users, which are all capable of performing IBFD communication. We assume that the entire frequency band is partitioned into $K$ subcarriers each with bandwidth $\omega$. Furthermore, the set of users and subcarriers are denoted by $\mathcal{N}$ and $\mathcal{K}$, respectively. It is considered that all the subcarriers are perfectly orthogonal to one another and no inter-subcarrier interference exists. We further assume that a subcarrier is exclusively assigned for the communications of a single user in both UL and DL. The subcarrier allocation variable is denoted by $a_{n,k}$ where
\begin{equation}
a_{n,k} = \begin{cases}
1, & \text{if subcarrier $k$ is assigned to user \textit{n}}, \\
0, & \text{otherwise}.
\end{cases}
\end{equation}

Moreover, $h^\mathrm{dl}_{n,k}$ and $h^\mathrm{ul}_{n,k}$ represent the channel coefficient of user $n$ in subcarrier $k$ in DL and UL, respectively. The DL SINR of user $n$ in subcarrier $k$ is defined as:
\begin{equation}
\gamma_{n,k}^\mathrm{dl}(\mathbf{p}_n,~\mathbf{q}_n) = \frac{p_{n,k} h^\mathrm{dl}_{n,k}}{q_{n,k} \Delta_\mathrm{u} + \sigma^2},
\end{equation}
where $\sigma^2$ and $\Delta_\mathrm{u}$ denote the noise density and SI-cancellation factor of user devices, respectively, and $p_{n,k}$ and $q_{n,k}$ are the DL and UL transmit power of user $n$ in subchannel $k$, in that order. Furthermore, we have $\mathbf{p}_{n}=[p_{n,1},...,p_{n,k}]$ and  $\mathbf{q}_{n}=[q_{n,1},...,q_{n,k}]$.

Since subcarrier $k$ is used for communications of user $n$ in both directions, UL and DL signals will interfere with one another, which results in SI. In equation (2), $q_{n,k} \Delta_\mathrm{u}$,  is the term that represents this residual SI in DL. Similarly, we define the UL SINR of user $n$ in subcarrier $k$ as:
\begin{equation}
\gamma_{n,k}^\mathrm{ul}(\mathbf{p}_n,~\mathbf{q}_n) = \frac{ q_{n,k} h^\mathrm{ul}_{n,k}}{p_{n,k} \Delta_\mathrm{bs} + \sigma^2},
\end{equation}
with $\Delta_\mathrm{bs}$ denoting SI-cancellation factor of MBS.

The data rate of user $n$ in subcarrier $k$ in DL and UL are:
\begin{equation}\label{rate}
r_{n,k}^\mathrm{dl} (\mathbf{a}_n,~\mathbf{p}_n,~\mathbf{q}_n) = a_{n,k}~\omega\log_2{(1+\gamma_{n,k}^\mathrm{dl}(\mathbf{p}_n,~\mathbf{q}_n))},
\end{equation}
and
\begin{equation}\label{rate}
r_{n,k}^\mathrm{ul} (\mathbf{a}_n,~\mathbf{p}_n,~\mathbf{q}_n) = a_{n,k}~\omega\log_2{(1+\gamma_{n,k}^\mathrm{ul}(\mathbf{p}_n,~\mathbf{q}_n))},
\end{equation}
respectively, where $\mathbf{a}_{n}=[a_{n,1},...,a_{n,k}]$. Accordingly, the total data rate of user $n$ in DL is:
\begin{equation}\label{totalrateuser}
R_n^\mathrm{dl} (\mathbf{a}_n,~\mathbf{p}_n,~\mathbf{q}_n) =\sum_{k\in \mathcal{K}}{r_{n,k}^\mathrm{dl}(\mathbf{a}_n,~\mathbf{p}_n,~\mathbf{q}_n)}.
\end{equation}

Similar to (6), the total data rate of user $n$ in UL, denoted by $R_n^\mathrm{ul}(\mathbf{a}_n,~\mathbf{p}_n,~\mathbf{q}_n)$, is obtained as $
R_n^\mathrm{ul} (\mathbf{a}_n,~\mathbf{p}_n,~\mathbf{q}_n) =\sum_{k\in \mathcal{K}}{r_{n,k}^\mathrm{ul}(\mathbf{a}_n,~\mathbf{p}_n,~\mathbf{q}_n)}.$
The total throughput of system is given by
\begin{equation}\label{R^T}
R^T(\mathbf{a},~\mathbf{p},~\mathbf{q}) = \sum_{n \in \mathcal{N}} \big(R_{n}^\mathrm{ul}(\mathbf{a}_n,~\mathbf{p}_n,~\mathbf{q}_n)+R_{n}^\mathrm{dl}(\mathbf{a}_n,~\mathbf{p}_n,~\mathbf{q}_n)\big),
\end{equation}
where $\mathbf{p}=\text{vec}[\mathbf{p}_{1},...,\mathbf{p}_{n}]$,~$\mathbf{q}=\text{vec}[\mathbf{q}_{1},...,\mathbf{q}_{n}]$, and $\mathbf{a}=\text{vec}[\mathbf{a}_{1},...,\mathbf{a}_{n}]$.

To compute the total energy consumption of network, we use the following energy consumption model in which both transmit power consumption and circuit energy consumption of devices are taken into account, and there are coefficients that represent the efficiency of power amplifiers in network devices
\begin{equation}\label{energy}
\begin{aligned}
E^T(\mathbf{a},~\mathbf{p},~\mathbf{q}) = &\sum_{n \in N}\sum_{k\in \mathcal{K}} a_{n,k}(\frac{1}{\kappa}p_{n,k}+\frac{1}{\psi}q_{n,k})\\
&+ N P^\mathrm{u}_\mathrm{c}+P_c^\mathrm{MBS}.
\end{aligned}
\end{equation}
In (\ref{energy}), $P^\mathrm{u}_\mathrm{c}$ and $P^\mathrm{MBS}_\mathrm{c}$ denote the circuit energy consumption of user device and MBS, respectively, and $\kappa$ and $\psi$ are power amplifier efficiency in MBS and user device, in that order.
Let us define EE as the ratio of system throughput to the corresponding network energy consumption, and denote it by $\eta_{EE}(\mathbf{a},~\mathbf{p},~\mathbf{q})$, where
\begin{equation}
\eta_{EE}(\mathbf{a},~\mathbf{p},~\mathbf{q}) = \frac{R^T(\mathbf{a},~\mathbf{p},~\mathbf{q})}{E^T(\mathbf{a},~\mathbf{p},~\mathbf{q})}.
\end{equation}
Moreover,is defined as follows
\begin{equation}
	 \eta_{SE}(\mathbf{a},~\mathbf{p},~\mathbf{q}) = \frac{R^T(\mathbf{a},~\mathbf{p},~\mathbf{q})}{W},
	 \end{equation}
where $W$ denotes the total bandwidth.

\section{Problem Statement}

The problem of joint subcarrier and power allocation for maximizing EE and SE under QoS and maximum transmit power constraints, is formally stated as:
\begin{equation}\label{first}
\begin{aligned}
&\max_{\mathbf{a},~\mathbf{p},~\mathbf{q}} \eta_{EE}(\mathbf{a},~\mathbf{p},~\mathbf{q})\\
&\max_{\mathbf{a},~\mathbf{p},~\mathbf{q}}\eta_{SE}(\mathbf{a},~\mathbf{p},~\mathbf{q})\\
\text{s.t.}~&C_1: \sum_{n \in \mathcal{N}}\sum_{k\in \mathcal{K}} a_{n,k} p_{n,k}\leq p_\mathrm{max},\\
&C_2: \sum_{k\in \mathcal{K}} a_{n,k} q_{n,k} \leq p^\mathrm{u}_\mathrm{max},~\forall n\in \mathcal{N},\\
&C_3: R_n^\mathrm{dl}(\mathbf{a}_n,~\mathbf{p}_n,~\mathbf{q}_n) \geq R_\mathrm{min}^\mathrm{dl},~ \forall n\in \mathcal{N},\\
&C_4: R_n^\mathrm{ul}(\mathbf{a}_n,~\mathbf{p}_n,~\mathbf{q}_n) \geq R_\mathrm{min}^\mathrm{ul},~ \forall n\in \mathcal{N},\\
&C_5: \sum_{n \in \mathcal{N}} a_{n,k}\leq 1,~ \forall k\in \mathcal{K},\\
&C_6: a_{n,k}\in \{0,1\},~ \forall n\in\mathcal{N}, \forall k \in \mathcal{K}.
\end{aligned}
\end{equation}

In the MOOP (\ref{first}), constraints $C_1$ and $C_2$ are related to  transmit power feasibility. Constraint $C_1$ indicates that the total transmit power of MBS should not exceed its maximum threshold which is denoted by $p_\mathrm{max}$, and $C_2$ restricts users maximum transmit power to $p_\mathrm{max}^\mathrm{u}$. In constraints $C_3$ and $C_4$, a minimum rate requirement is guaranteed for each user in DL,~$R_\mathrm{min}^\mathrm{dl}$, and UL, $R_\mathrm{min}^\mathrm{ul}$, respectively. Constraint $C_5$ indicates that each subchannel can be allocated to at most one user and in $C_6$, the binary nature of subchannel allocation variable is implied.

Due to the binary subchannel allocation variables and the interference included in rate function, problem (\ref{first}) is a mixed-integer non-linear program (MINLP) which is generally difficult to solve. 
In the following section, we restate problem (11) as an equivalent MOOP, whose purpose is to maximize system throughput and minimize energy consumption, simultaneously.

\section{Proposed Solution}

As given in (9), EE is the ratio of throughput and energy consumption and note that $R^{T}=\eta_{SE}W$,~therefore, we can write $\eta_{EE}=\frac{\eta_{SE}W}{E^{T}}$.~It is straightforward to deduct that maximization of $\eta_{EE}(\mathbf{a},~\mathbf{p},~\mathbf{q})$ is equivalent to maximizing $R^T(\mathbf{a},~\mathbf{p},~\mathbf{q})$ while minimizing $E^T(\mathbf{a},~\mathbf{p},~\mathbf{q})$, simultaneously~\cite{14}. To this end, we reformulate (\ref{first}) as an equivalent MOOP that is given in (\ref{second}):

\begin{equation}\label{second}
\begin{aligned}
&f_1: \min_{\mathbf{a},~\mathbf{p},~\mathbf{q}}~E^{T}(\mathbf{a},~\mathbf{p},~\mathbf{q})\\
&f_2: \max_{\mathbf{a},~\mathbf{p},~\mathbf{q}}~R^{T}(\mathbf{a},~\mathbf{p},~\mathbf{q})\\
&\text{s.t.} ~~C_1 - C_6.
\end{aligned}
\end{equation}

The first objective of the optimization problem (\ref{second}), $f_1$, is to minimize system energy consumption, and the second one, $f_2$, is to maximize system's throughput and its constraint set is the same as that of (\ref{first}).

Even though the MOOP (\ref{second}) contains two competing objective functions, we can still find a solution for it that satisfies the predefined conditions of Pareto optimal fronts,~here,~we employ $\epsilon$-constraint method~\cite{11} by keeping $f_{1}$ as as the primary objective function and moving $f_{2}$ to the constraint set The new optimization problem would be:
\begin{equation}\label{third}
\begin{aligned}
&f_1:~\min_{\mathbf{a},~\mathbf{p},~\mathbf{q}}E^{T}(\mathbf{a},~\mathbf{p},~\mathbf{q})\\
\text{s.t.}~
& C_0: R^T(\mathbf{a},~\mathbf{p},~\mathbf{q}) \geq \epsilon,\\
&C_1 - C_6.
\end{aligned}
\end{equation}

Due to the multiplication of variables, $\mathbf{a}$, $\mathbf{p}$, and $\mathbf{q}$, it is still non-convex and thus challenging to address. Furthermore, $C_0$, requires the total throughput of system to be greater that $\epsilon$. It is obvious that the feasibility of (\ref{third}) as well as the the closeness of its solution to the solution of problem (\ref{first}), greatly depend on the value of $\epsilon$. This fact turns $\epsilon$ into a sensitive parameter, whose value should be carefully estimated. Moreover, we are still faced with the same challenges in dealing with the non-convex constraint set of (\ref{first}).

In order to address the non-convex optimization problem (\ref{third}), we first deal with the problem of variables multiplication in constraints $C_1$ and $C_2$. In the left-hand side of these constraints, it is implied that if subchannel $k$ is not allocated to user $n$ ($a_{n,k} =0 $), the transmit power of this user over $k$ should be zero in both UL and DL ($p_{n,k}=q_{n,k}=0$). Based on this explanation, we can restate $C_1$ and $C_2$ as follows:
\begin{align}
&C'_1: \sum_{n \in \mathcal{N}}\sum_{k\in \mathcal{K}} p_{n,k}\leq p_\mathrm{max},\\
&C''_1: p_{n,k} \leq a_{n,k}p_\mathrm{max},\\
&C'_2: \sum_{k\in \mathcal{K}} q_{n,k}\leq p^\mathrm{u}_\mathrm{max},\\
&C''_2: q_{n,k} \leq a_{n,k}p^\mathrm{u}_\mathrm{max}.
\end{align}




Another challenge in solving (\ref{third}) is the integer subcarrier allocation variable, $a_{n,k}$. This binary variable turns (\ref{third}) into a MINLP, which is difficult to solve in an acceptable timespan. To address this issue, we take an approach similar to \cite{9,20}, and replace constraint $C_6$ with the following inequalities:
\begin{align}
&C'_6: 0 \leq a_{n,k} \leq 1,~\forall n\in \mathcal{N}, \forall k \in \mathcal{K},\\
&C''_6: \sum_{n\in \mathcal{N}}\sum_{k\in \mathcal{K}}a_{n,k} -a^2_{n,k}\leq 0.
\end{align}

As the last step in converting the constraint set of problem (\ref{third}) into a convex set, we should deal with the non-convex rate functions,  $R_n^\mathrm{ul}(\mathbf{p}_n,~\mathbf{q}_n)$ and $R_n^\mathrm{dl}(\mathbf{p}_n,~\mathbf{q}_n)$. Let us rewrite $R_n^\mathrm{dl}(\mathbf{p}_n,~\mathbf{q}_n)$ as follows:
\begin{equation}
R_n^\mathrm{dl}(\mathbf{p}_n,~\mathbf{q}_n) = f^\mathrm{dl}_{n}(\mathbf{p}_{n},~\mathbf{q}_{n}) - g^\mathrm{dl}_{n}(\mathbf{p}_{n},~\mathbf{q}_{n}),
\end{equation}
where
\begin{equation}
f^\mathrm{dl}_{n}(\mathbf {p}_{n},~\mathbf{q}_{n}) =\sum_{k \in {\mathcal{K}}}  \log_2 (p_{n,k}h^\mathrm{dl}_{n,k}+q_{n,k}\Delta_\mathrm{u}+\sigma^2),
\end{equation}
and,
\begin{equation}
g^\mathrm{dl}_{n}(\mathbf{q}_{n}) = \sum_{k \in {\mathcal{K}}} \log_2 (q_{n,k}\Delta_\mathrm{u}+\sigma^2).
\end{equation}

The equality given in (20) consists of two concave functions, $f^\mathrm{dl}_{n}(\mathbf{p}_{n},~\mathbf{q}_{n})$ and $g^\mathrm{dl}_{n}(\mathbf{q}_{n})$. However, the subtraction of these concave functions is not necessarily convex. To tackle this issue, we find a convex approximation for $R_n^\mathrm{dl}(\mathbf{p}_{n}, \mathbf{q}_{n})$ by using majorization minimization (MM) method~\cite{18}. In this method, a series of surrogate functions are constructed that approximate the originally non-convex function. Here, we use Taylor approximation for constructing our surrogate function. To do so, in iteration number $t$ we will have:

\begin{equation}\label{tildeg}
\tilde{g}^\mathrm{dl}_{n}(\mathbf{q}_{n}) = g_{n}^\mathrm{dl}(\mathbf{q}_{n}^{t-1})+\nabla_\mathbf{q}g^T(\mathbf{q}^{t-1}). (\mathbf{q}-\mathbf{q}^{t-1}).
\end{equation}

Based on (\ref{tildeg}), we define the convex approximation of DL rate function, ${R}_n^\mathrm{dl}(\textbf{p}_n,~\textbf{q}_n)$, as:
\begin{equation}
\tilde{R}_n^\mathrm{dl}(\mathbf{p}_n,~\mathbf{q}_n) = f^\mathrm{dl}_{n}(\mathbf{p}_{n},~\mathbf{q}_{n}) - \tilde{g}^\mathrm{dl}_{n}(\mathbf{q}_{n}).
\end{equation}

Since $\tilde{g}^\mathrm{dl}_{n}(\mathbf{q}_{n})$ is an affine function and $f^\mathrm{dl}_{n}(\mathbf{p}_{n},~\mathbf{q}_{n})$ is convex, $\tilde{R}_n^\mathrm{dl}(\mathbf{p}_{n},~\mathbf{q}_{n})$ is a convex approximation of ${R}_n^\mathrm{dl}(\mathbf{p}_{n},~\mathbf{q}_{n})$.
Similarly, the approximate UL data rate would be:
\begin{equation}
\tilde{R}_n^\mathrm{ul}(\mathbf{p}_n,~\mathbf{q}_n) = f^\mathrm{ul}_{n}(\mathbf{p}_{n},~\mathbf{q}_{n}) - \tilde{g}^\mathrm{ul}_{n}(\mathbf{p}_{n}),
\end{equation}
where
\begin{equation}
f^\mathrm{ul}_{n}(\mathbf {p}_{n},~\mathbf{q}_{n}) =\sum_{k \in {\mathcal{K}}}  \log_2 (q_{n,k}h^\mathrm{ul}_{n,k}+p_{n,k}\Delta_\mathrm{bs}+\sigma^2),
\end{equation}
\begin{equation}
\tilde{g}^\mathrm{ul}_{n}(\mathbf{p}_{n}) = g_{n}^\mathrm{ul}(\mathbf{p}_{n}^{t-1})+\nabla_\mathbf{p}g^T(\mathbf{p}^{t-1}). (\mathbf{p}-\mathbf{p}^{t-1}),
\end{equation}
and
\begin{equation}
g^\mathrm{ul}_{n}(\mathbf{p}_{n}) = \sum_{k \in {\mathcal{K}}} \log_2 (p_{n,k}\Delta_\mathrm{bs}+\sigma^2).
\end{equation}

Regarding the above transformations, we define the approximate total data rate of system as:
\begin{equation}
\tilde{R}^{T}(\mathbf{p},~\mathbf{q}) = \sum_{n \in \mathcal{N}} (\tilde{R}_n^\mathrm{dl}(\mathbf{p}_n,~\mathbf{q}_n)+ \tilde{R}_n^\mathrm{ul}(\mathbf{p}_n,~\mathbf{q}_n)).
\end{equation}

After these modifications, the resulting optimization problem would be:
\begin{equation}\label{last-1}
\begin{aligned}
&\min_{\mathbf{a},~\mathbf{p},~\mathbf{q}}E^{T}(\mathbf{p},~\mathbf{q})\\
\text{s.t.~}
&C_0: \tilde{R}^{T}(\mathbf{p},~\mathbf{q})\geq \epsilon\\
&C'_1,~C''_1,~C'_2,~C''_2,~C_5,~C'_6,~C''_6\\ 
&C_3: \tilde{R}_n^{dl}(\mathbf{p},~\mathbf{q}) \geq R_\mathrm{min}^\mathrm{dl},~ \forall n\in \mathcal{N},\\
&C_4: \tilde{R}_n^{ul}(\mathbf{p},~\mathbf{q}) \geq R_\mathrm{min}^\mathrm{ul},~ \forall n\in \mathcal{N},\\
\end{aligned}
\end{equation}

 Since constraint $C_6''$ is concave and greater or equal to zero, (\ref{last-1}) does not comply with the standard form of a convex optimization problem. To deal with this issue and facilitate the solution design, we remove constraint $C_6''$ from the constraint set of problem (\ref{last-1}) and add it as a penalty function\footnote{In fact, $\lambda$ acts as a penalty factor to penalize the objective function when $a_{n,k}$ is not binary value.}, with a weighting factor denoted by $\lambda$, to the objective function. After this modification, we will have:
\begin{equation}\label{last+1}
\begin{aligned}
&\min_{\mathbf{a},~\mathbf{p},~\mathbf{q}} ~E^T(\mathbf{p},\mathbf{q})+\lambda\Big(\sum_{n \in \mathcal{N}}\sum_{k\in \mathcal{K}}\big(a_{n,k}-a_{n,k}^2\big)\Big)\\
\text{s.t.:}~& C_0,~C'_1,~C''_1,~C'_2,~C''_2,~C_3,~C_4,~C_5,~C'_6.
\end{aligned}
\end{equation}
 
 \noindent
 \textbf{Remark:}
   It can be easily demonstrated that the optimization problem (\ref{last+1})~is equivalent to (\ref{last-1}). For more details, refer to \cite{9},~\cite{20}.

 To tackle the non-convexity of objective function in the above problem, we first rewrite the objective function as:
\begin{equation}
e( \mathbf{a},~\mathbf{p},~\mathbf{q}) = e_1( \mathbf{a},~\mathbf{p},~\mathbf{q}) - \lambda e_2(\mathbf{a}),
\end{equation}
where $e_1( \mathbf{a},~\mathbf{p},~\mathbf{q})=E^T(\mathbf{p},~\mathbf{q})+\lambda \big(\sum_{n \in \mathcal{N}}\sum_{k\in \mathcal{K}}a_{n,k}\big)$ and
$e_2(\mathbf{a})=\big(\sum_{n \in \mathcal{N}}\sum_{k\in \mathcal{K}}a_{n,k}^{2}\big)$. Now we use a similar approach that was previously explained for approximation of the rate functions, and estimate $e_2(\mathbf{a})$ as:
\begin{equation}\label{four}
\begin{aligned}
\tilde{e_2}(\mathbf{\mathbf{a}})= e_2(\mathbf{a}^{t-1})+\nabla_{\mathbf{a}} e_2^T(\mathbf{a}^{t-1})(\mathbf{a}-\mathbf{a}^{t-1}).
\end{aligned}
\end{equation}

Eventually, the resulting convex optimization problem would be:
\begin{equation}\label{last}
\begin{aligned}
&\min_{\mathbf{a},~\mathbf{p},~\mathbf{q}}e_1(\mathbf{a},~\mathbf{p},~\mathbf{q}) - \lambda \tilde{e}_2(\mathbf{a})\\
& \text{s.t.}~C_0,~C'_1,~C''_1,~C'_2,~C''_2,~C_3,~C_4,~C_5,~C'_6.
\end{aligned}
\end{equation}

The optimization problem (\ref{last}) is a convex optimization problem. In order to solve this problem and obtain a locally optimal solution for problem (\ref{third}), here we employ the difference of convex functions (DC) programming \cite{19}.


\noindent
\textbf{Proposition}:
	 The solution obtained for (\ref{last}) by incorporating DC approximation at the end of each iteration, is a locally optimal solution for the original problem (\ref{first}).

It should be noted that, constraint $C_0$ in optimization problem (\ref{last}) asserts that the total throughput of network, $\tilde{R}^T(\mathbf{a},~\mathbf{p},~\mathbf{q})$, should be greater than or equal to $\epsilon$. To further clarify the impact of $\epsilon$ on the optimization problem (\ref{last}), let us consider the three following cases:
\begin{itemize}
	\item[i.]if $\epsilon = 0$, optimization problem (\ref{last}) would turn into the problem of minimizing system's energy consumption.
	\item[ii.] if $\epsilon = R_\mathrm{max}$, assuming $R_\mathrm{max}$ is the maximum system throughput, the solution obtained for (\ref{last}) would be the solution of network throughput maximization problem.
	\item[iii.]  if $\epsilon \geq R_\mathrm{max}$, the optimization problem (\ref{last}) would be infeasible.
\end{itemize}

Regarding the above cases, it can be easily deducted that the optimization problem (\ref{last}) and its obtained solution are very sensitive to the value of $\epsilon$ and through this parameter, a trade-off between system's throughput and aggregate energy consumption can be derived.

From cases (i) and (ii), we can perceive that the maximum value that $\epsilon$ can take without making (\ref{last}) infeasible is $R_\mathrm{max}$. Since $R_{\max}$ is maximum system throughput, we can obtain its value by solving the following optimization problem:
\begin{equation}\label{R_max}
\begin{aligned}
&\max_{\mathbf{a},~\mathbf{p},~\mathbf{q}}\tilde{R}^{T}(\mathbf{p},~\mathbf{q})-\lambda\Big( \big(\sum_{n \in \mathcal{N}}\sum_{k\in \mathcal{K}}a_{n,k}\big)-\tilde{e}_2(\mathbf{a})\Big)\\
 \text{s.t.}~&C'_1,~C''_1,~C'_2,~C''_2,~C_3,~C_4,~C_5,~C'_6,
\end{aligned}
\end{equation}
which is in fact the optimization problem of maximizing system's throughput.
By solving problem (\ref{R_max}), the maximum value of $\epsilon$ would be determined.

Since different values of $\epsilon$ result in different trade-offs between system's throughput and energy consumption,~we should find a value for $\epsilon$ that corresponds to the maximum $\tilde{R}^T(\mathbf{a},~\mathbf{p},~\mathbf{q})$ to $E^T(\mathbf{a},~\mathbf{p},~\mathbf{q})$ ratio. To find this specific value of $\epsilon$, we use the equality below:
\begin{equation}\label{eps}
\epsilon = \delta R_\mathrm{max},
\end{equation}
where $\delta$ is a positive value in the range of ($0,1$]. Depending on the value of $\delta$, the ratio between system's throughput and energy consumption varies; however, for a specific $\delta$ this ratio reaches a maximum value.
\begin{table}
  \centering
\caption{simulation parameters}
\label{Simulation Parameters}
\begin{tabular}{|c|c|}\hline
{\bf Parameter} & {\bf Value} \\ \hline \hline
{Cell radius} & $350$ m \\
Number of users & $10$\\
 Number of sub-channels & \{$16,32,64$\} \\
Noise power ($\sigma^{2}$) & {$-120$} dBm \\
{Path-loss exponent } & {$3$} \\
 {$P_c^\textnormal{u}$} & {$0.1$ W} \\
 {$P_m^\textnormal{MBS}$} & {$1$ W} \\
 {$\kappa$}&{38\%}\\
 {$\psi$}&{20\%}\\
$p_{\textrm{max}}$ & {$42$ dBm} \\
$p_{\textrm{max}}^{u}$ & {$23$ dBm} \\
$R_{\textrm{min}}^{dl}$ & $5$ bps/Hz \\
$R_{\textrm{min}}^{ul}$ & $2.5$ bps/Hz \\
$\lambda$ & $10^{6}$\\
\hline
\end{tabular}
\end{table}

\section{Simulation Results}

We evaluate the performance of our proposed resource allocation algorithm through extensive simulations. In our simulations, we consider a macrocell with radius $350$ m and $K=64$ subchannels. We further assume that there are $N=10$ users that communicate in IBFD mode. The channel gain between a transmitter and a receiver is calculated using independent and identically distributed Rayleigh flat fading and the figures shown in this section are obtained by calculating the average of results over different realizations of path loss as well as multipath fading. Without loose of generality we assume that BS and users' SI-cancellation factors are the same and $\Delta_\mathrm{u}$ = $\Delta_\mathrm{bs}$ = $\Delta$ = -90 dB.~The rest of the simulation parameters are given in Table I.

We first examine the effect of SI-cancellation factor, $\Delta$, on energy efficiency of IBFD networks. In Fig. 1, system energy efficiency vs. $\delta$  for different values of $\Delta$ is presented. We also draw a comparison between EE of IBFD communications and that of HD in Fig. 1. For HD case, we assume that half of the existing subchannels are reserved for DL and the other half for UL communications, exclusively. Due to the concavity of rate functions in HD communications (because of the absence of interference), we use Dinklebach method to obtain the solution of joint subchannel and power allocation for EE maximization problem in a HD single cell network.

As observed in Fig. 1, by decreasing $\Delta$, system EE would increase. This is due to the fact that lower values of $\Delta$ correspond to lower SI and thus higher EE. Furthermore, in each IBFD case, for a specific $\delta$, EE reaches its peak and then decreases. However, the value of $\delta$ for which the maximum EE is obtained, varies from one case to another. For instance, when $\Delta = -90$ dB, for $\delta = 0.7$ the maximum EE is achieved while for $\Delta = -70$ dB, system EE peaks in $\delta = 0.6$. This observation can be explained by considering the amount of data rate that a user can attain by consuming a unit of energy. When $\Delta = -90$ dB, because of the lower SI, user would be able to achieve a notable data rate, even while transmitting with a nominal transmit power. In this case, since the substantial growth in system throughput is worth the slight increase in system power consumption, the $\delta$ for which the maximum EE is attained leans toward higher values. In contrast, when SI intensity is high, the value of $\delta$ corresponding to the maximum EE would get closer to lower values of $\delta$. Another important observation in Fig. 1  is the superiority of IBFD communications' performance compared to HD. Note that as $\delta$ gets closer to its optimal value (in peaks), the EE achieved using IBFD becomes higher than EE of HD. This improved performance is the results of the higher flexibility of spectrum usage in IBFD communications.

 \begin{figure}
\includegraphics[width=9cm,height=5cm]{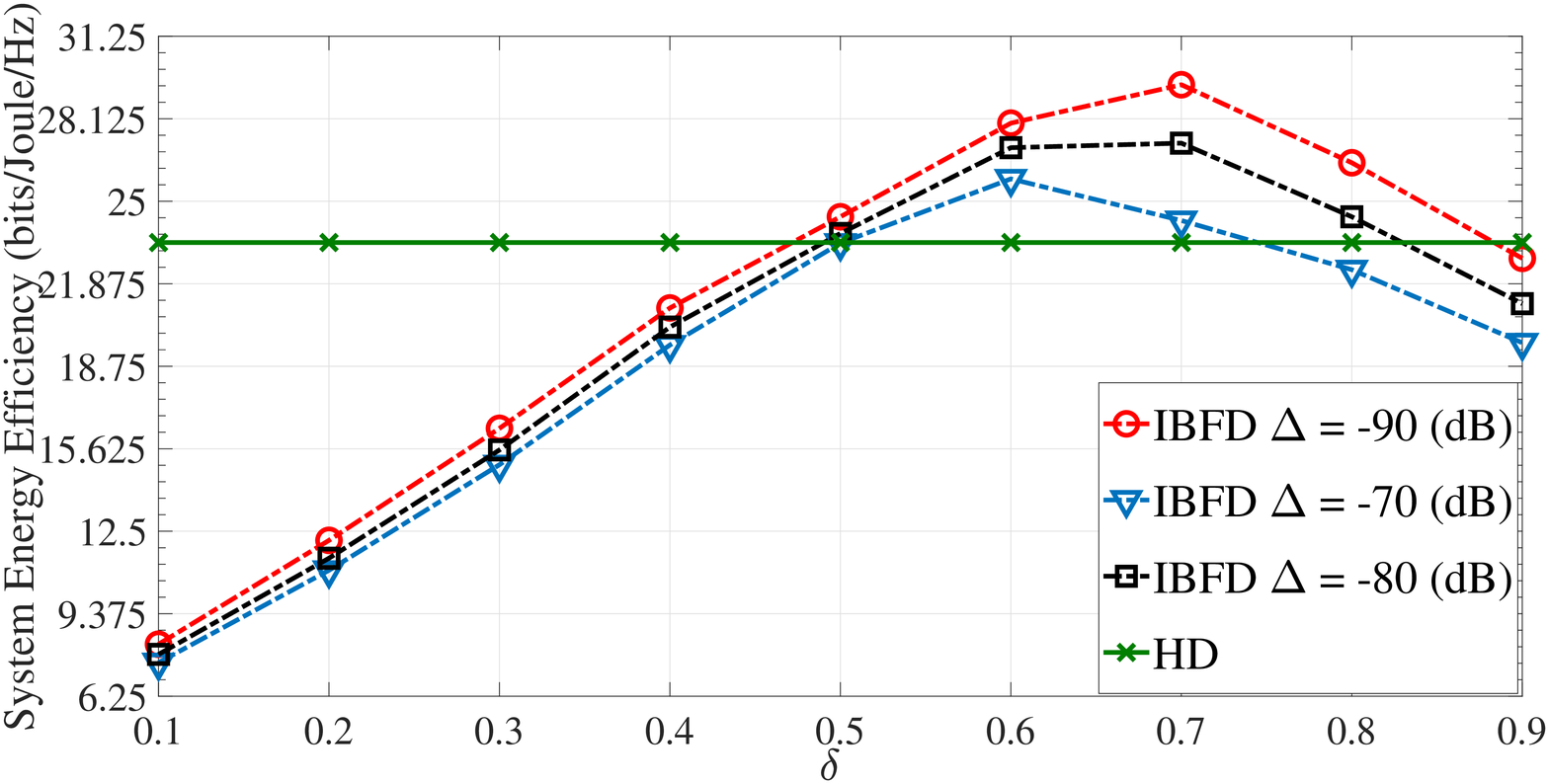}
	\caption{System energy efficiency vs. $\delta$ for different $\Delta$}
\end{figure}
\begin{figure}[t]
	\centering
\includegraphics[width=9cm,height=5cm]{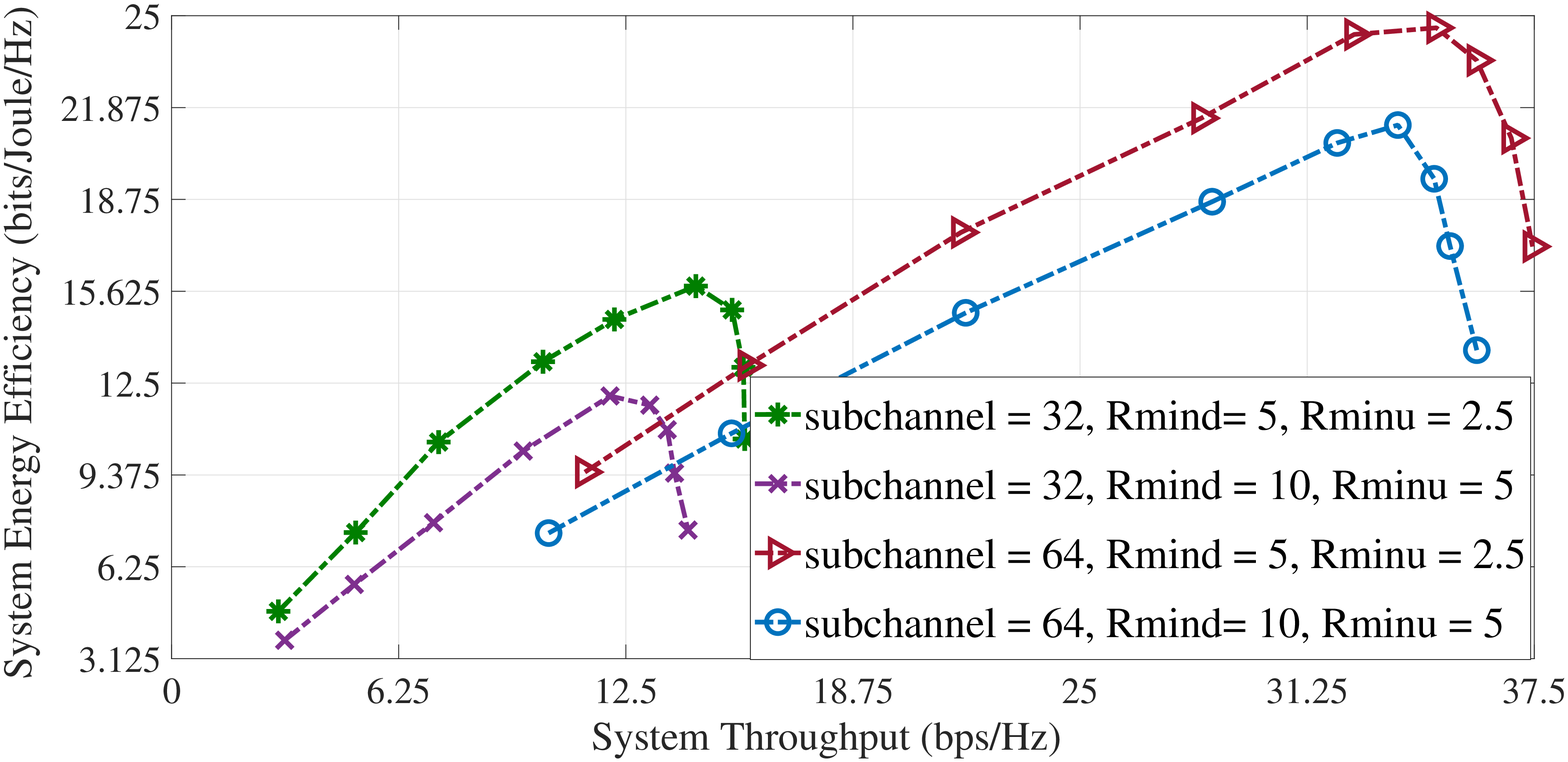}
	\caption{EE vs. SE}
\end{figure}
\begin{figure}
\includegraphics[width=9cm,height=5cm]{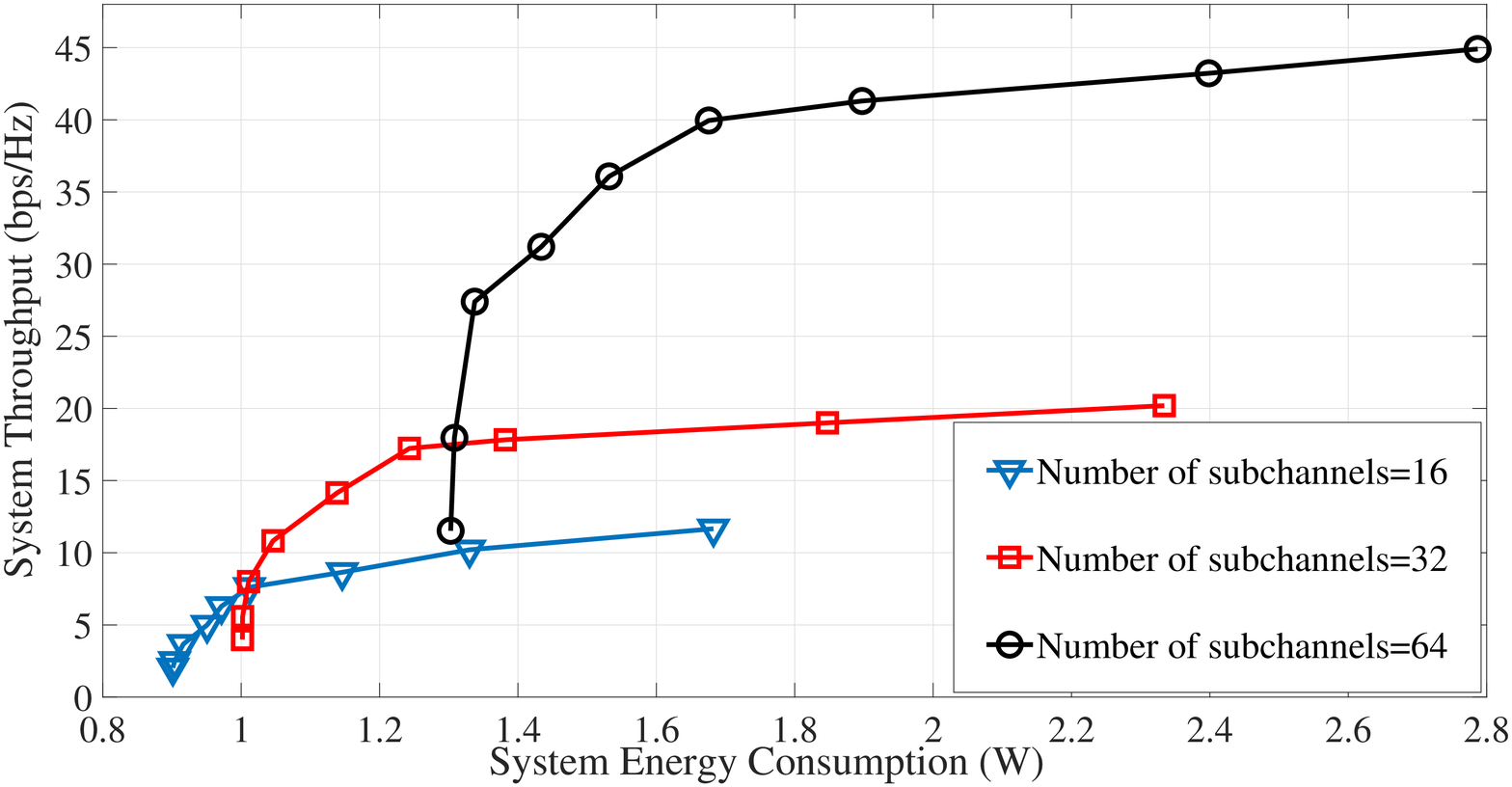}
	\caption{System throughput vs. system energy consumption}
\end{figure}

In Fig. 2, the trend of system EE with respect to system
throughput is illustrated. In this figure, we notice that as the
throughput of network increases, EE steadily grows and then
sharply decreases with it. In fact, system throughput is by
itself a function of system transmit power, thus any increase
in throughput also means more energy consumption. Since EE
is the ratio of system throughput to energy consumption, when
the cost of rise in system total data rate, which is the amount
of energy consumed in system, becomes far greater than the
gain we achieve by it, system EE starts to decline with any
increase in network throughput.

In Fig. 3, we can see the rate of change in system throughput
with respect to variations of system energy consumption for
different number of subchannels. It is evident that for
all cases, any small increase in system energy consumption
results in a considerable increase in system throughput. This
growth becomes more notable as the number of subchannels
in the system increases, which is due to the existence of more
available subchannels that can be exploited for improving
system throughput. On the other hand, from one specific value
of energy consumption onward, no matter how radical the
change is in energy consumption, the rate of change in system throughput becomes quite subtle. For instance, when $K = 64$,
until $E^{T} (\mathbf{p},~\mathbf{q})\approx 1.7~W$, the increase in system throughput
is quite remarkable, however, when $E^{T} (\mathbf{p},~\mathbf{q})$ exceeds this
value, this rate of change slows down and system throughput
almost converges. This observation can be explained by taking
the amount of interference that sending with high transmit
power (and thus consuming more energy) would cause, into
account. In fact, even though high transmit power can result
in higher SNR (therefore higher data rate), users' SINR does
not comply with this rule. Meaning, increase in transmit power
may prevail the negative effect of the interference caused by it
up to a point, however, from that point onward, the amount of
interference that such high transmit power engenders, lessens
the rate of increase in system throughput.

\section{Conclusion}

In this paper we investigated the problem of joint subchanel assignment and power control to strike a balance between EE and SE in an OFDMA network with IBFD communications. This problem was formulated as a MOOP in order to maximize system throughput and minimize aggregate power consumption, simultaneously. To obtain all the Pareto fronts in the aforementioned problem, we have used the $\epsilon$-constraint method. Furthermore, in order to tackle the non-convexity of the constraint set, a majorization minimization approach has been used for approximating the non-convex rate functions and a penalty function was introduced to handle the binary subchannel allocation variables. The effectiveness of IBFD communications, as well as the capability of our proposed solution in improving EE as well as SE of network was demonstrated through simulations.

\end{document}